\documentclass[conference]{IEEEtran}
\IEEEoverridecommandlockouts

\usepackage{array,multirow}
\usepackage{booktabs} 
\usepackage{graphicx}
\usepackage{caption}
\usepackage{subfig}
\usepackage{amssymb}
\usepackage{threeparttable}
\usepackage{booktabs}
\usepackage{cite}
\usepackage{bm}
\usepackage{bbm}
\usepackage{amsmath,amssymb,amsfonts}
\usepackage{algorithmic}
\usepackage{graphicx}
\usepackage{mathtools}
\usepackage{textcomp}
\usepackage{xcolor}
\usepackage[hyphens]{url}

\usepackage[capitalize]{cleveref}
\crefname{section}{Sec.}{Secs.}

\def\BibTeX{{\rm B\kern-.05em{\sc i\kern-.025em b}\kern-.08em
    T\kern-.1667em\lower.7ex\hbox{E}\kern-.125emX}}

\definecolor{color0}{HTML}{FFD700}
\definecolor{color1}{HTML}{EA5F94}
\definecolor{color2}{HTML}{9D02D7}
\definecolor{color3}{HTML}{0000FF}

\definecolor{darkblue}{HTML}{00429D}
\definecolor{darkgreen}{HTML}{005c00}
\definecolor{gold}{HTML}{D4AF37}
\definecolor{darkred}{HTML}{910000}
\usepackage{tikz}
\usepackage{pgfplots}
\usepackage{tikzscale}
\usepackage{tikz-qtree}
\pgfplotsset{compat=1.16}
\pgfplotsset{every axis/.append style={
                    label style={font=\scriptsize},
                    tick label style={font=\scriptsize},
                    legend style={font=\scriptsize}
                    }}
\usetikzlibrary{plotmarks,shapes,patterns,decorations.pathreplacing,backgrounds,calc,arrows,arrows.meta,spy,matrix,shadows,trees,positioning,fit}
\usepgfplotslibrary{patchplots,groupplots}

\tikzstyle{startstop} = [rectangle, rounded corners, minimum width=2cm, minimum height=0.5cm,text centered, draw=black]
\tikzstyle{io} = [trapezium, trapezium left angle=70, trapezium right angle=110, minimum width=3cm, minimum height=1cm, text centered, draw=black]
\tikzstyle{process} = [rectangle, minimum width=2cm, minimum height=0.5cm, text centered, draw=black, alignb=center]
\tikzstyle{decision} = [ellipse, minimum width=2cm, minimum height=1cm, text centered, draw=black]
\tikzstyle{arrow} = [thick,<->,>=stealth]
\tikzstyle{line} = [thick,>=stealth]
\tikzstyle{darrow} = [thick,<->,>=stealth,dashed]
\tikzstyle{sarrow} = [thick,->,>=stealth]
\tikzstyle{larrow} = [line width=0.1mm,dashdotted,->,>=stealth]

\newlength\fheight
\newlength\fwidth

\makeatletter
\def\endthebibliography{%
	\def\@noitemerr{\@latex@warning{Empty `thebibliography' environment}}%
	\endlist
}
\makeatother

\usepackage[acronyms,nonumberlist,nopostdot,nomain,nogroupskip]{glossaries}

\newacronym{drl}{DRL}{Deep Reinforcement Learning}
\newacronym{dqn}{DQN}{Deep Q-Network}
\newacronym{iot}{IoT}{Internet of Things}
\newacronym{wot}{WoT}{Web of Things}
\newacronym{tdd}{TDD}{Thing Description Directory}
\newacronym{dt}{DT}{Digital Twin}
\newacronym{td}{TD}{Thing Description}
\newacronym{ai}{AI}{Artificial Intelligence}
\newacronym{qos}{QoS}{Quality of Service}
\newacronym{tt}{TT}{Thermal Twin}
\newacronym{rt}{RT}{Room Twin}
\newacronym{mse}{MSE}{Mean Square Error}
\newacronym{mdp}{MDP}{Markov Decision Process}

\def \mwidth{0.24\linewidth}
\def \twidth{0.28\linewidth}

\def \fwidth{0.9\columnwidth}
\def \fheight {0.6\columnwidth}

\def \theight {0.25\linewidth}

\begin{document}

\title{A Web of Things Architecture for Digital Twin Creation and Model-Based Reinforcement Control}

\author{\IEEEauthorblockN{Luca Bedogni}
\IEEEauthorblockA{Department of Physics, Mathematics and Informatics\\
University of Modena and Reggio Emilia\\
Via G. Campi 213/B, 41125 Modena, Italy\\
\texttt{luca.bedogni@unimore.it}}\and
\IEEEauthorblockN{Federico Chiariotti}
\IEEEauthorblockA{Department of Information Engineering\\
University of Padova\\
Via G. Gradenigo 6/B, 35131 Padova, Italy\\
\texttt{chiariot@dei.unipd.it}}
\thanks{This work was partly supported by the Italian Ministry of University and Research under the PNRR ``SoE Young Researchers'' grant for project REDIAL.}}

\maketitle

\begin{abstract}
\gls{iot} devices are available in a multitude of scenarios, and provide constant, contextual data which can be leveraged to automatically reconfigure and optimize smart environments. To realize this vision, \gls{ai} and deep learning techniques are usually employed, however they need large quantity of data which is often not feasible in \gls{iot} scenarios. \glspl{dt} have recently emerged as an effective way to replicate physical entities in the digital domain, to allow for simulation and testing of models and services. In this paper, we present a novel architecture based on the emerging \gls{wot} standard, which provides a \gls{dt} of a smart environment and applies \gls{drl} techniques on real time data. We implement our system in a real deployment, and test it along with a legacy system. Our findings show that the benefits of having a digital twin, specifically for \gls{drl} models, allow for faster convergence and finer tuning.
\end{abstract}


\glsresetall

\section{Introduction}
\label{sec:introduction}
\glsresetall

The \gls{iot} has transformed many aspects of daily lives, as modern, intelligent devices are now available in a plethora of different use cases. This vast amount of information, together with the capability of \gls{iot} devices to interact with the real world, opens up several possibilities for the automatic reconfiguration and control of systems based on \gls{ai} and deep learning techniques, without human intervention in the loop \cite{Montori2018M2MSurvey}.

The \gls{drl} revolution, which started in 2015 with the first functioning \gls{dqn} model~\cite{mnih2015human}, has led to a flourishing of proposed solutions in various fields, including the \gls{iot}. In the years since, \gls{drl} models have proven their value in a number of applications, from complex board games to cellular network optimization. However, some inherent challenges still limit their use in practical scenarios~\cite{dulac2021challenges}: while these models can often overcome traditional approaches after convergence, the training phase requires some exploration of the state and action spaces, leading to suboptimal outcomes during training. 

This training phase might be extremely long in some applications, as training even a modestly sized \gls{dqn} often requires millions of steps. Moreover, several models may be a viable solution for the same task, hence understanding the one which fits the best is not trivial. This can effectively prevent the use of \gls{drl} in many domotic and industrial \gls{iot} applications~\cite{wang2020reinforcement}, whose control loops are relatively slow, as the training would require weeks or even months of highly suboptimal control. On the other hand, purely passive training (i.e., training the model on experience samples that were obtained while using a policy not controlled by the learning agent) has known convergence issues~\cite{ostrovski2021difficulty}: if a traditional control policy is used during training, avoiding highly damaging outcomes, the \gls{dqn} might never learn a good policy.

A solution to this issue is given by \emph{model-based} \gls{drl}~\cite{kidambi2020morel}: in this paradigm, domain knowledge about the task of the learning agent is used to create a model of the environment, which can be used as a simulation environment for offline virtual training. The agent is then free to make mistakes and learn from experience, without actually causing any damage in the real world, and the model can be sped up to make the duration of the training process independent from the timing of the task, as the only limit to the virtual environment is the available processing speed. Naturally, the model itself is not reality, and the agent might need an online training phase in the real world after reaching convergence in the simulated environment, but the difficulty of transfer learning from the virtual environment to the real one is much lower than learning the policy from scratch, and more importantly, extremely damaging actions are already associated to very low values, and a directed exploration policy will never select them, achieving a much smaller optimality gap during the online training phase.

Enabling simulation of counterfactual scenarios and different control policies is one of the key aims of another paradigm that is rapidly attracting interest in the engineering community, the \emph{\gls{dt}}~\cite{boschert2016digital}. \glspl{dt} include sensory, simulation, and management aspects of objects and processes, effectively bridging the physical and digital worlds~\cite{twin5gacia}: a twin can be used to monitor the state of its physical counterpart, by connecting it directly to sensors, or to simulate the effects of choices and actions, using it as a dynamic model disconnected from physical reality. In fact, a \gls{dt} can provide the virtual environment for model-based \gls{drl}, acting as a sandbox to train the learning agent in without affecting the real environment. At the same time, the twin can be refined with updated data from the real sensors, more accurately matching the physical system it represents; the \gls{drl} agent can then be easily retrained on the improved sensor data, incorporating the new model with the experience on the actual physical system.

In this work, we describe a general-purpose \gls{wot} architecture that can seamlessly support the creation and twinning of \glspl{dt} and the model-based training of \gls{drl} agents in a virtual environment composed by one or more \glspl{dt}. The architecture can interface directly with sensors and actuators through \gls{wot} standard protocols, as well as providing a middleware layer that can integrate legacy systems and proprietary \gls{iot} frameworks, such as most domotic systems, into the architecture. We also provide a Smart Home use case, in which \gls{drl} is used to control the heating system of a room, considering the number of people occupying it as well as associated energy efficiency. The \gls{dt} and \gls{drl} agent are connected to a real deployment, using data from the installed sensors and providing notifications to the user with advice on thermostat settings. As more data becomes available over time, the \gls{dt} for each room is automatically updated, exploiting more complex models when their training becomes possible, and the deployed \gls{drl} agent is retrained using the new models, providing a self-improving temperature control with a seamless occupant experience.

The rest of this paper is organized as follows: first, we provide a review of the relevant literature in Sec.~\ref{sec:sota}. We then describe our architecture and the model-based \gls{drl} training procedure in Sec.~\ref{sec:arch}, and we present our domotic use case in Sec.~\ref{sec:drlusecase}. Finally, Sec.~\ref{sec:conclusions} concludes the paper and presents some possible avenues of future work.

\section{Related Work}\label{sec:sota}
The \gls{iot} has revolutionized many aspects of daily lives, becoming pervasive and present in several different scenarios. To fully leverage the \gls{iot}'s potential, Deep Learning or Machine Learning techniques are often employed to extract patterns from the vast amount of data that sensors and other \gls{iot} devices produce. The \gls{iot} presents many challenges, one of which is certainly the possibility for them to cooperate even if made by different vendors and adopting different communication protocols \cite{Montori2018M2MSurvey}. This is one of the reasons behind the increasing interest over the \gls{wot} standard, which provides a homogeneous representation of devices and of operations which can be performed on them \cite{Ragget2015WoT,Zeng2011WoTSurvey}, while also offering \gls{qos} levels depending on the application requirements, and security mechanisms to protect the data and the devices. 

However, even being able to seamlessly interconnect devices does not fully unleash the possibilities offered by the \gls{iot}. What is still missing is the possibility for devices to automatically understand what services can spark from the devices and the data which is available in their networks. In other words, what is missing is the interpretation of the data, which goes beyond pure classification \cite{Montori2018Classification}, to go towards automatic service composition built by the device from the data itself \cite{Bedogni2022CCNC}. There have been many contributions in this domain, most of which focused on a specific aspect of the issue, such as how to discover devices using the same protocol and exchanging similar information \cite{Georgakopoulos2015Discovery}, and how to leverage ontologies to provide semantics to the data, hence to be able to provide a shared sense to it \cite{Agarwal2016Ontology} \cite{Seydoux2016Ontology}. 

There are fewer works which address the autonomous creation of services from the devices, and more generally how different devices can form novel services without human intervention \cite{Monster2017Causality}. Apart from understanding the nature of the data, devices also need to be able to perform different actions on the network to change its behavior and recognize what actions may bring the system toward a better state \cite{Lake2017BuildingMachinesThink}. In other words, the network has to automatically understand what is the causality of actions on specific variables of interest, which requires to perform several experiments and possibly have a long history of data, which is not always feasible in an \gls{iot} scenario. There have been some advances in this domain~\cite{Lippi2021SenseOfAgency}, but there are still no organic solutions that can also deal with all the challenges that deployment in real scenarios brings~\cite{Monster2017Causality,Agadakos2018Causality}.

During the past decade, another revolution in computing and \gls{ai} took place, in parallel to the development of \gls{iot} systems: reinforcement learning, whose practical applications had been mostly limited to board games and parametric optimization of hand-designed algorithms, was successfully integrated with deep neural networks~\cite{mnih2015human}, taking a significant leap in terms of performance and capabilities. Over the past few years, \gls{drl} has become a staple solution in many fields, including robotics, automated factory management, and communication networks. However, \gls{drl} solutions for real systems have two significant problems, which stem from the model-free nature of the learning procedure: firstly, they require a long training process, which must include samples of highly suboptimal actions in order to fully explore the solution space, and secondly, counterfactuals and alternative options are hard to evaluate~\cite{madumal2020explainable}.

Domotics, and the control of slow processes such as building heating and climate control, are an application field that is particularly vulnerable to long training times~\cite{wang2020reinforcement}: while training a \gls{drl} agent in faster environments, with sub-second time steps, can take a few hours, training time for agents on processes whose actions are separated by minutes or hours can be measured in months. Furthermore, the robustness of the learned solution needs to be verified thoroughly, as the consequences of failures may range from annoying to dire. 

In these cases, virtual pre-training before deployment is a common technique~\cite{lei2022practical}. Complex simulators or simple black-box models can provide a safe environment fro the \gls{drl} agent to make mistakes~\cite{di2021deep}, transferring the acquired knowledge to the real world.
However, most works using virtual environments for training designed them \emph{ad hoc}~\cite{pinto2022transfer}, requiring extensive human intervention, and the quality of the simulation environment is crucial for effective transfer learning~\cite{schreiber2021monitoring}.

Causal reasoning and the creation of models of the environment, either in advance or during training and operation, are ways to mitigate these issues: modeling the effects of actions, and whether the available actions affect control performance~\cite{seitzer2021causal}, can significantly improve \gls{drl} training. The concept of a \gls{dt}, which uses real data from the environment to build a model, can be extremely useful: by gradually improving the environment model, and retraining the \gls{drl} agent if needed, the \gls{dt} paradigm can aid the virtual environment design process~\cite{matulis2021robot}, providing the basic building blocks for automated data-driven training~\cite{xia2021digital}.

This work combines ideas from the \gls{dt} and \gls{wot} literature in a single architecture, which can easily integrate multiple \glspl{dt} from different types of sensors and automatically retrain a \gls{drl} agent for complex control task in a virtual environment generated by a selection of \glspl{dt}.

\section{Architecture}\label{sec:arch}
In this section, we describe the building blocks of our framework. We leverage the \gls{wot} model, which enables a standardized interface and communication process for smart devices belonging to a network. We focus our efforts on a Smart Home scenario, but our architecture can be easily extended and adapted to other environments as well.

In Sec.~\ref{sec:wot}, we describe our framework, which is based on a \gls{wot} network through which devices are able to communicate. The training procedure for model-based \gls{drl} agents is then described in Sec.~\ref{sec:drltrain}, including the twinning of the digital twin model to the selected sensory set and the training of the agent in the virtual environment provided by the twin. We then describe the inclusion of legacy components in Sec.~\ref{sec:legacy}, to extend our vision and encompass other devices which may be available through other networks or systems. 

\begin{figure}
    \centering
    \begin{tikzpicture}[every text node part/.style={align=center}]

\node[draw,rounded corners,draw=darkblue,fill=darkblue,fill opacity=0.3,text opacity=1,minimum height=1cm,text width=2cm,minimum width=2cm] (tdd) at (0,0) {\scriptsize Thing Description Directory};

\node[draw,rounded corners,draw=darkgreen,fill=darkgreen,fill opacity=0.3,text opacity=1,minimum height=1cm,text width=2cm,minimum width=2cm] (eng) at (3,0) {\scriptsize Engine};

\node[draw,rounded corners,draw=gold,fill=gold,fill opacity=0.3,text opacity=1,minimum height=1cm,text width=2cm,minimum width=2cm] (wot) at (-3,0) {\scriptsize WoT sensors and actuators};

\node[draw,rounded corners,draw=darkgreen,fill=darkgreen,fill opacity=0.3,text opacity=1,minimum height=1cm,text width=2cm,minimum width=2cm] (mw) at (0,1.5) {\scriptsize Middleware};

\node[draw,rounded corners,draw=darkblue,fill=darkblue,fill opacity=0.3,text opacity=1,minimum height=1cm,text width=2cm,minimum width=2cm] (leg1) at (3.2,1.7) {};
\node[draw,rounded corners,draw=darkblue,fill=white,text opacity=1,minimum height=1cm,text width=2cm,minimum width=2cm] (legb1) at (3.1,1.6) {};
\node[draw,rounded corners,draw=darkblue,fill=darkblue,fill opacity=0.3,text opacity=1,minimum height=1cm,text width=2cm,minimum width=2cm] (leg2) at (3.1,1.6) {};
\node[draw,rounded corners,draw=darkblue,fill=white,text opacity=1,minimum height=1cm,text width=2cm,minimum width=2cm] (legb2) at (3,1.5) {};
\node[draw,rounded corners,draw=darkblue,fill=darkblue,fill opacity=0.3,text opacity=1,minimum height=1cm,text width=2cm,minimum width=2cm] (leg) at (3,1.5) {\scriptsize Legacy systems};

\foreach \n in {0,...,2}
    \node[draw,rounded corners,draw=gray,fill=gray,fill opacity=0.3,text opacity=1,minimum height=0.5cm,text width=0.75cm,minimum width=0.75cm] (m\n) at (0.9+1.2*\n,3.25) {\scriptsize DT \n};

\foreach \n in {0,...,2}
    \node[draw,rounded corners,draw=darkred,fill=darkred,fill opacity=0.3,text opacity=1,minimum height=0.5cm,text width=0.75cm,minimum width=0.75cm] (a\n) at (-0.9-1.2*\n,3.25) {\scriptsize DRL \n};

\node[](p1) at (0.5,0.75){};

\draw[->] ([xshift=0.1cm]tdd.north) -- ([xshift=0.1cm]mw.south);
\draw[<-] ([xshift=-0.1cm]tdd.north) -- ([xshift=-0.1cm]mw.south);
\draw[->] (eng.west) -- (tdd.east);
\draw[->] ([yshift=0.1cm]wot.east) -- ([yshift=0.1cm]tdd.west);
\draw[<-] ([yshift=-0.1cm]wot.east) -- ([yshift=-0.1cm]tdd.west);
\draw[-] (leg.south) |- (0.5,0.75);
\draw[->] (0.5,0.75) -- ([xshift=0.5cm]tdd.north);
\draw[->] (mw.east) -- (leg.west);

\draw[-] (mw.north) -- (0,2.5);
\foreach \n in {0,...,2}
    \draw[->] (0,2.5) -| (m\n.south);
\foreach \n in {0,...,2}
    \draw[->] (0,2.5) -| (a\n.south);
\foreach \n in {0,...,2}
    \draw[-] (-4,3.75) -| (a\n.north);
\draw[-] (-4,3.75) |- (-0.5,0.75);
\draw[->] (-0.5,0.75) -| ([xshift=-0.5cm]tdd.north);

\end{tikzpicture}
    \caption{Web of Things architecture.}\vspace{-0.5cm}
    \label{fig:architecture}
\end{figure}
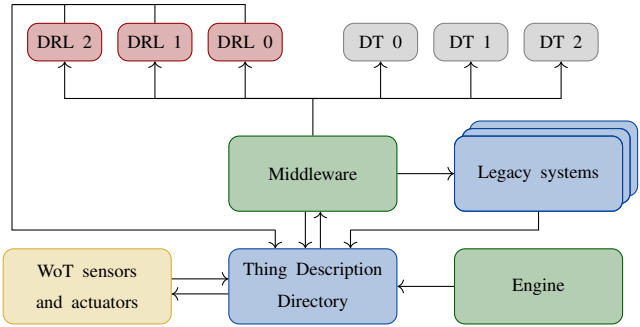

\begin{figure*}
\centering
    \input{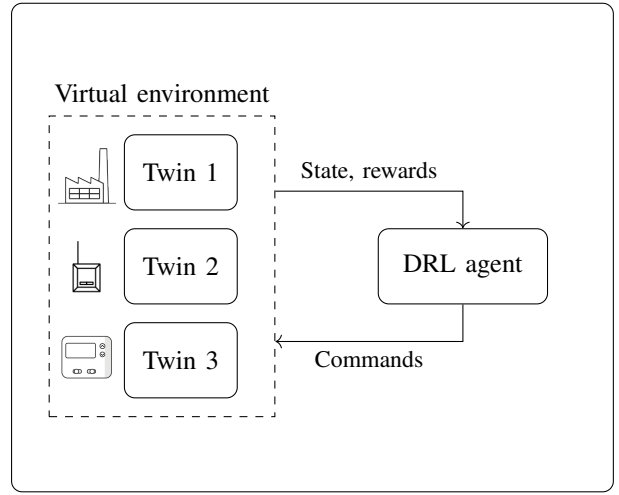}
    \caption{Diagram of the two stages of a model-based \gls{drl} training in the \gls{wot} platform.}\vspace{-0.5cm}
    \label{fig:drltraining}
\end{figure*}

\subsection{WoT Network}
\label{sec:wot}
Our architecture extends the simpler version presented in \cite{Bedogni2022CCNC} and mainly builds on the \gls{wot} paradigm, which enables devices to provide their representation in terms of a \gls{td}, which advertises their own capabilities to other devices willing to interact with them. Fig.~\ref{fig:architecture} highlights the different components which build our architecture and their interactions.

On the lower left cprner of the figure, we see sensors and actuators, which are built following the WoT standard: they provide a minimalist HTTP server which serves the requests, and also provide their \gls{td} whenever their root URL is visited. 

Upon joining the network, devices are also requested to join a \gls{tdd}, which stores all the \glspl{td} from devices in the network. Using a \gls{tdd} has numerous advantages, such as the possibility to perform queries on it to retrieve specific groups of devices, or to gather all devices providing the same kind of data and so on. Moreover, it allows to keep track of all the devices which are available in our system. The \gls{tdd} also provides an interface to notify other software components of devices joining the network. To this interface we subscribe two other components, which are the Middleware and the Engine.

The Middleware is in charge of recognizing devices joining the network and instantiate an appropriate model for interpreting the values by performing two main operations:
\begin{itemize}
    \item It monitors the \gls{tdd} for devices which have joined the network, analyzes their \gls{td} and instantiates an appropriate model which can handle such type of data. Moreover, it can also instantiate an appropriate \gls{dt} which can handle that device.
    \item It connects to legacy systems which may be already present in the smart environment, and provides a \gls{wot} interface to them. These newly created \gls{wot} devices will then be registered to the \gls{tdd}, so that other devices in the system can leverage their data and can possibly interact with them.
\end{itemize}
Overall, the Middleware is in charge of maintaining updated the modules and \gls{dt} components according to the devices active in the system. Therefore, it also periodically checks whether registered devices are still alive and removes the models or \glspl{dt} whenever such devices left the network.

Finally, the Engine is in charge of optimizing the system by considering at all the Things registered in the TDD and their capabilities, as well as monitoring the agent rewards. 

\subsection{Model-Based Training}\label{sec:drltrain}

The proposed architecture includes a two-step procedure for training model-based \gls{drl} agents, as shown in Fig.~\ref{fig:drltraining}: the basic concept is to exploit the Digital Twin paradigm to provide a virtual environment that is safe for an agent to explore, without any impacts on the actual physical system.

\begin{enumerate}
\item \emph{Digital Twin training and data gathering:} Firstly, the system needs to gather sensory observation and create a digital twin of the observed environment. In our architecture, the \gls{tdd} may include different sensors and actuators, belonging to different logical and physical environments and tasks: for example, two temperature sensors might be in different parts of a building or plant, and two sensors in the same room might have different logical functions, which make them functionally independent (e.g., light and temperature sensors). The system administrator then needs to select the set of sensors to gather in a single \gls{dt} by querying the \gls{tdd}. Once the model is instantiated, the Middleware layer automatically subscribes it to the sensors fitting the specified \gls{td}, starting the twinning process. The platform is entirely agnostic to the training method used for the model, which could use either statistical or learning-based methods: since the architecture operates over standardized queries and responses, the \glspl{dt} can be implemented as a black box, even using different programming languages. During this stage, the actuators in the environment must be operated using a legacy policy, which may be defined by an algorithm or directly by the user, as shown in Fig.~\ref{fig:drltraining}.
\item \emph{Virtual model-based \gls{drl} training:} After the \glspl{dt} have reached the required accuracy, a \gls{drl} agent can be trained in a virtual environment without affecting the real system. The twins can be used in a predictive fashion, applying the results of the agent's choices and simulating the real system. The virtual environment may include more than one twin, representing different subsystems that are affected by the agent's actions and concur in determining the system reward. Naturally, the choice of which \glspl{dt} must be included, as well as which actuators can be controlled by the \gls{drl} agent, is made by the system administrator. 
\end{enumerate}

From a \gls{drl} point of view, the model-based training is robust, as long as the \glspl{dt} represent the effects of actions accurately. In any case, a short online training phase in the real environment is still possible, with a small cost in terms of system efficiency. During this phase, a directed exploration algorithm will still avoid any catastrophic failure modes, as the Agent will have learned that they lead to significant penalties in the virtual environment. On the other hand, from the \gls{wot} perspective, the Middleware just needs to connect the Agent to the proper components: during the virtual training, observations of the environment are provided by the \glspl{dt}, while during real operation, observations from the sensors are pushed directly to the Agent. In the same way, its commands are only directed to the \glspl{dt} during the virtual training, but are routed to the real Actuators in the operational phase.

\subsection{Legacy Components}
\label{sec:legacy}
In order to integrate our architecture into an existing system, the Middleware also includes the possibility to interact with legacy systems already available in the smart environment. To accomplish this, it is also possible to design specific components directly in the Middleware to interconnect to other systems exposing data or services.

A note on this is that those specific components, developed to connect to legacy systems, have to be also deployed as a Thing themselves. This allows for them to be registered in the \gls{tdd} upon connecting to the legacy system, so that other devices can leverage the data exposed by them, or connect to the offered services. Clearly, the precise definition of one or more legacy components has to be made according to the scenario and to already existing deployments, as it enables our system to integrate and cooperate with the others.

\section{Use Case: Heating System DRL Control}
\label{sec:drlusecase}
In this section we describe a prototype implementation of our system, and we detail how we implemented all the software components described in Section \ref{sec:arch}. We consider a smart home scenario, in which different sensors can provide environmental data in order to schedule activities and optimize the planning. We focus our efforts on deploying our system aside a legacy system which manages different sensors and actuators around a house, which are not compliant to the \gls{wot} standard. This scenario can be found in many modern homes, which leverage data from off-the-shelf devices which also provide proprietary software solutions to manage them. We base our analysis on the climate operations of the house, which means that our aim is to optimize the temperature of different rooms accounting for the room occupancy and for the inside and outside temperature. Specifically, we deploy our system in a real house deployment with 2 floors, three rooms and four inhabitants which provide heterogeneity in objectives and room occupancy, which are:
\begin{itemize}
    \item A living room open space, which also includes the kitchen;
    \item A bathroom;
    \item A master bedroom shared by two people.
\end{itemize}
The living room is generally occupied during the morning, at lunchtime and during the evenings, but it may also get sporadic occupation throughout the entire day in case people work from home or at lunch and dinner times. A bathroom typically has a flat occupancy during the day, meaning that people can use it at different times without a specific routine. On the other hand, the bedroom instead has zero occupancy during the day, and full occupancy during the night, although bedtime and wake ups can have relatively small variations on different days of the week. We will detail the parameters we took into account below.

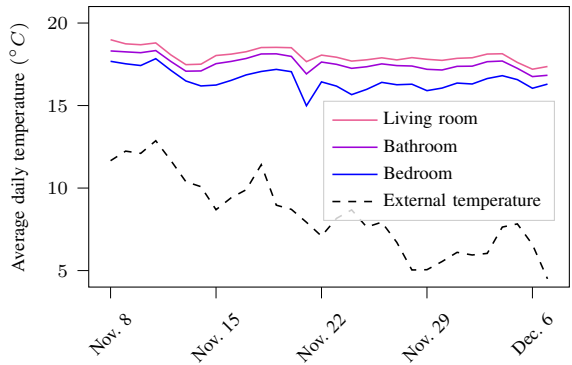
\begin{figure}
\centering
\pgfplotsset{scaled x ticks=false}
\begin{tikzpicture}

\definecolor{darkgray176}{RGB}{176,176,176}
\definecolor{lightgray204}{RGB}{204,204,204}

\begin{axis}[
width=\fwidth,
height=\fheight,
legend cell align={left},
legend style={
  fill opacity=0.8,
  draw opacity=1,
  text opacity=1,
  at={(0.97,0.45)},
  anchor=east,
  draw=lightgray204
},
tick align=outside,
tick pos=left,
x grid style={darkgray176},
xmin=19302.55, xmax=19334.45,
xtick style={color=black},
xtick={19304,19311,19318,19325,19332},
xticklabels={Nov. 8, Nov. 15, Nov. 22, Nov. 29, Dec. 6},
xticklabel style={rotate=45.0},
y grid style={darkgray176},
ylabel={Average daily temperature $(^\circ C)$},
ymin=4, ymax=21,
ytick style={color=black}
]
\addplot [semithick, color1]
table {%
19304 18.9890023148148
19305 18.7397853835979
19306 18.6846878306878
19307 18.794424382716
19308 18.0699652777778
19309 17.4781944444444
19310 17.5064083994709
19311 18.0321805555556
19312 18.1157943121693
19313 18.2606348104056
19314 18.5154954805996
19315 18.5259363425926
19316 18.5041150793651
19317 17.6603167438272
19318 18.0559748677249
19319 17.9275204475309
19320 17.6934795524691
19321 17.7731903659612
19322 17.8932546296296
19323 17.7653474426808
19324 17.9030661926808
19325 17.802931547619
19326 17.7383075396825
19327 17.860630787037
19328 17.8952868165785
19329 18.1205432098765
19330 18.1395027006173
19331 17.6122569444444
19332 17.2019236111111
19333 17.3658383838384
};
\addlegendentry{Living room}
\addplot [semithick, color2]
table {%
19304 18.3086284722222
19305 18.2537916666667
19306 18.2028506944444
19307 18.3324027777778
19308 17.67359375
19309 17.0821875
19310 17.0968472222222
19311 17.5382534722222
19312 17.6736631944444
19313 17.8523402777778
19314 18.1260892857143
19315 18.1339461805556
19316 17.9774479166667
19317 16.9163680555556
19318 17.6364375
19319 17.4969548611111
19320 17.254921875
19321 17.3520138888889
19322 17.525625
19323 17.4203240740741
19324 17.3953732638889
19325 17.1960396825397
19326 17.1547048611111
19327 17.37653125
19328 17.3865104166667
19329 17.6578298611111
19330 17.6975347222222
19331 17.2548090277778
19332 16.7545138888889
19333 16.8339393939394
};
\addlegendentry{Bathroom}
\addplot [semithick, color3]
table {%
19304 17.6806944444444
19305 17.5306597222222
19306 17.4249375
19307 17.8398055555556
19308 17.1357986111111
19309 16.4889236111111
19310 16.1878125
19311 16.2424097222222
19312 16.5246527777778
19313 16.8571875
19314 17.0648660714286
19315 17.1917465277778
19316 17.05125
19317 14.9884444444444
19318 16.4341666666667
19319 16.1771736111111
19320 15.6612604166667
19321 15.9815625
19322 16.4050694444444
19323 16.2605439814815
19324 16.2933680555556
19325 15.9029166666667
19326 16.0589583333333
19327 16.3622222222222
19328 16.3061458333333
19329 16.636
19330 16.8105833333333
19331 16.5703472222222
19332 16.0475
19333 16.2987121212121
};
\addlegendentry{Bedroom}
\addplot [semithick, black, dashed]
table {%
19304 11.6583402777778
19305 12.238125
19306 12.1025694444444
19307 12.867
19308 11.6716666666667
19309 10.3840277777778
19310 10.0844444444444
19311 8.68788194444444
19312 9.41013888888889
19313 9.89839583333333
19314 11.4155486111111
19315 8.96320138888889
19316 8.70936111111111
19317 7.92786111111111
19318 7.08850694444444
19319 8.16579861111111
19320 8.67783035714286
19321 7.63615972222222
19322 7.93232638888889
19323 6.70788888888889
19324 5.03633333333333
19325 5.053125
19326 5.54669444444444
19327 6.10528472222222
19328 5.94510416666667
19329 6.04061111111111
19330 7.63625
19331 7.82802083333333
19332 6.57909722222222
19333 4.49848484848485
};
\addlegendentry{External temperature}
\end{axis}

\end{tikzpicture}
    \caption{Temperature dataset used for our experiments.}\vspace{-0.5cm}
    \label{fig:dataset}
\end{figure}
In Figure \ref{fig:dataset}, we also show the temperature dataset we have collected over 40 days. This data has been obtained through the legacy system sensors which were already deployed in the house in November-December 2022. The black line represents the outdoor temperature and shows a significant drop, from roughly 12$^{\circ}C$ to around 5$^{\circ}C$. The other three lines show the temperature inside the three rooms considered as part of this study. While the absolute values differ, their trends are similar and also reflect the current configuration of the already existing heating system. The heating in the three rooms is controlled by a classical thermostat system, in which the inhabitants set the desired temperature in different part of the day and the heating system is switched on to match that temperature. The schedule varies greatly for the three rooms: the master bedroom is only heated at night, the living room only during the day, while the bathroom is heated in the morning and in the evening. Nevertheless, they show a similar trend as heat spreads through the whole house.

\subsection{Implementation}
The house was already running a legacy system provided by Home Assistant\footnote{https://www.home-assistant.io/}, which is a personal smart home hub providing integration with several off-the-shelf devices. Home Assistant allows users to interconnect components offered by different vendors, and provides a standardized domain and naming schemes to obtain data and perform actions. We describe at first the implementation of the Legacy component connecting the middleware to Home Assistant, and we then highlight how it is integrated into our architecture.

Home Assistant provides data with different APIs. For our purpose, we performed the integration through the data exposed in InfluxDB, a popular time-series database which is integrated within Home Assistant and which stores data from the devices managed by Home Assistant. Since our aim is to be \gls{wot} compliant to make devices able to register in the \gls{tdd}, we developed a software component as a Thing, which performs the following:
\begin{itemize}
    \item It connects to InfluxDB and reads all the devices providing a variable of interest. In our case, since we are focusing on optimizing the temperature, the corresponding domain is \texttt{climate}, which is the group of devices in Home Assistant which can provide temperature, humidity and also perform operations such as switching on a cooling or heating system.
    \item For each discovered device, it provides a specific GET operation advertised in its thing description. For each of these methods in the \gls{td}, whenever they are requested it will translate the query to an appropriate InfluxDB query to retrieve the data and eventually provide it to the client requesting it. 
\end{itemize}
This software component is then registered in the \gls{tdd}, so that other devices have seamless access to its data without requiring specific knowledge on the technical details of the connection.

\begin{figure}[t]
\centering
\subfloat[Node-RED integration into Home Assistant\label{fig:node-red}]{\includegraphics[width=.9\columnwidth]{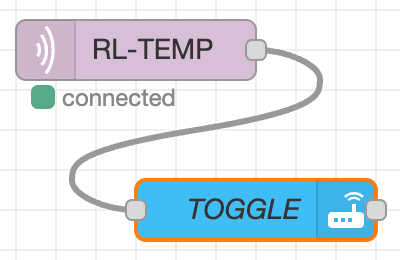}}\\
\subfloat[Dashboard visualization and state.\label{fig:dashboard}]{\includegraphics[width=.9\columnwidth]{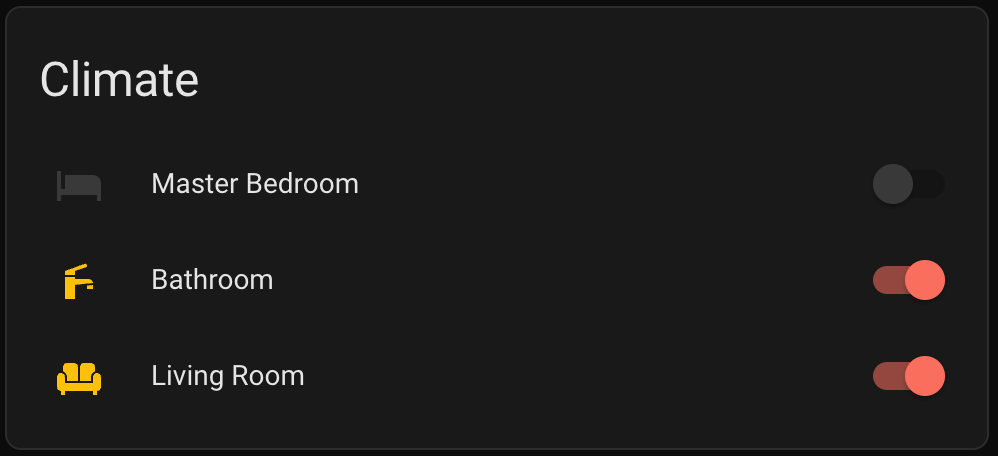}}%
\caption{Integration of the legacy component output inside the Home Assistant instance.}
\label{fig:legacy}
\end{figure}

The developed software component also provides bidirectional communication, meaning that it can also feedback to Home Assistant. This is realized by exposing a PUT method in the TD of the software, which allows clients to send data to the Thing, which then routes it back to Home Assistant. To realize this part, without loss of generality we have adopted the MQTT\footnote{MQTT originally stood for Message Queuing Telemetry Transport, but the acronym has officially not stood for anything since 2013.} protocol, by publishing data on the Home Assistant MQTT broker. Clearly, other protocols can be adopted depending on the scenario and on the specific needs. We show an example of this behavior in Fig.~\ref{fig:legacy}: Fig.~\ref{fig:node-red} shows the MQTT subscribe (RL-TEMP) node inside Node-RED running in Home Assistant. Figure \ref{fig:dashboard} shows instead the Home Assistant dashboard, with the three heating systems managed by our models and with their current statuses.

\subsection{Digital Twin Evaluation}
\label{sec:performance}
In this section, we show the performance evaluation of our model in a real deployment, leveraging a \gls{tt} which twins the temperature dynamics accounting for outside and inside temperatures, and a \gls{rt} which instead accounts for the rooms occupancy. 

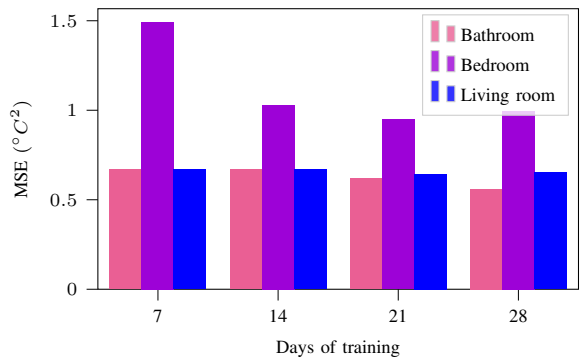
\begin{figure}
\centering
\begin{tikzpicture}

\definecolor{darkgray176}{RGB}{176,176,176}
\definecolor{darkslategray66}{RGB}{66,66,66}
\definecolor{lightgray204}{RGB}{204,204,204}

\begin{axis}[
width=\fwidth,
height=\fheight,
legend cell align={left},
legend style={fill opacity=0.8, draw opacity=1, text opacity=1, draw=lightgray204},
tick align=outside,
tick pos=left,
unbounded coords=jump,
x grid style={darkgray176},
xlabel={Days of training},
xmin=-0.5, xmax=3.5,
xtick style={color=black},
xtick={0,1,2,3},
xticklabels={7,14,21,28},
y grid style={darkgray176},
ylabel={MSE $(^\circ C^2)$},
ymin=0, ymax=1.56728535668426,
ytick style={color=black}
]
\draw[draw=none,fill=color1] (axis cs:-0.4,0) rectangle (axis cs:-0.133333333333333,0.669672240366031);
\addlegendimage{ybar,ybar legend,draw=none,fill=color1}
\addlegendentry{Bathroom}

\draw[draw=none,fill=color1] (axis cs:0.6,0) rectangle (axis cs:0.866666666666667,0.667728971523303);
\draw[draw=none,fill=color1] (axis cs:1.6,0) rectangle (axis cs:1.866666666666667,0.619474509961589);
\draw[draw=none,fill=color1] (axis cs:2.6,0) rectangle (axis cs:2.866666666666667,0.560623863804258);

\draw[draw=none,fill=color2] (axis cs:-0.133333333333333,0) rectangle (axis cs:0.13333333333333,1.49265272065168);
\addlegendimage{ybar,ybar legend,draw=none,fill=color2}
\addlegendentry{Bedroom}

\draw[draw=none,fill=color2] (axis cs:0.866666666666667,0) rectangle (axis cs:1.13333333333333,1.03052449824978);
\draw[draw=none,fill=color2] (axis cs:1.866666666666667,0) rectangle (axis cs:2.13333333333333,0.949308522470598);
\draw[draw=none,fill=color2] (axis cs:2.866666666666667,0) rectangle (axis cs:3.13333333333333,0.994868020294303);

\draw[draw=none,fill=color3] (axis cs:0.133333333333333,0) rectangle (axis cs:0.4,0.667962484086355);
\addlegendimage{ybar,ybar legend,draw=none,fill=color3}
\addlegendentry{Living room}

\draw[draw=none,fill=color3] (axis cs:1.133333333333333,0) rectangle (axis cs:1.4,0.667962326535708);
\draw[draw=none,fill=color3] (axis cs:2.133333333333333,0) rectangle (axis cs:2.4,0.644611060176846);
\draw[draw=none,fill=color3] (axis cs:3.133333333333333,0) rectangle (axis cs:3.4,0.652529490849849);

\end{axis}

\end{tikzpicture}
    \caption{MSE for the three considered rooms as a function of the amount of available data.}
    \label{fig:mse}
    \vspace{-0.5cm}
\end{figure}

The \gls{tt} was adapted from grey box RC-equivalent models in the relevant literature~\cite{leprince2022fifty}. We considered two measured parameters, the room's internal temperature $T_i$ and the ambient temperature outside the building $T_a$, which are available from the smart home sensors, along with a set of hidden variables which can help in tracking the thermal evolution of a room. The $T_iT_e$ model includes the temperature of the envelope (i.e., the building's external walls), and the $T_iT_h$ model includes the temperature of the heating system. The $T_iT_eT_h$ model includes both, and the $T_iT_eT_hR_{ia}$ includes a variable thermal resistance between the air in the room and the exterior of the building. Naturally, more complex models are more accurate, but require more data to avoid overfitting.

\begin{figure}
\centering
\begin{tikzpicture}

\definecolor{darkgray176}{RGB}{176,176,176}
\definecolor{darkslategray38}{RGB}{38,38,38}

\begin{axis}[
tick align=outside,
tick pos=left,
x grid style={darkgray176},
xlabel={Days of training},
xmin=0, xmax=4,
xtick style={color=black},
xtick={0.5,1.5,2.5,3.5},
xticklabels={7,14,21,28},
y dir=reverse,
y grid style={darkgray176},
ylabel={Room},
ymin=0, ymax=3,
ytick style={color=black},
ytick={0.5,1.5,2.5},
yticklabel style={rotate=90.0},
yticklabels={Bathroom,Bedroom,Living room}
]
\addplot graphics [includegraphics cmd=\pgfimage,xmin=0, xmax=4, ymin=3, ymax=0] {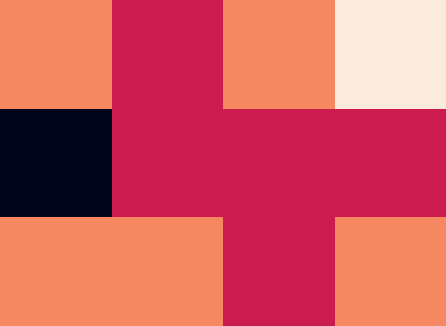};
\draw (axis cs:0.5,0.5) node[
  scale=0.8,
  text=white,
  rotate=0.0
]{$T_iT_eT_h$};
\draw (axis cs:1.5,0.5) node[
  scale=0.8,
  text=white,
  rotate=0.0
]{$T_iT_h$};
\draw (axis cs:2.5,0.5) node[
  scale=0.8,
  text=white,
  rotate=0.0
]{$T_iT_eT_h$};
\draw (axis cs:3.5,0.5) node[
  scale=0.8,
  text=darkslategray38,
  rotate=0.0
]{$T_iT_eT_hR_{ia}$};
\draw (axis cs:0.5,1.5) node[
  scale=0.8,
  text=white,
  rotate=0.0
]{$T_i$};
\draw (axis cs:1.5,1.5) node[
  scale=0.8,
  text=white,
  rotate=0.0
]{$T_iT_h$};
\draw (axis cs:2.5,1.5) node[
  scale=0.8,
  text=white,
  rotate=0.0
]{$T_iT_h$};
\draw (axis cs:3.5,1.5) node[
  scale=0.8,
  text=white,
  rotate=0.0
]{$T_iT_h$};
\draw (axis cs:0.5,2.5) node[
  scale=0.8,
  text=white,
  rotate=0.0
]{$T_iT_eT_h$};
\draw (axis cs:1.5,2.5) node[
  scale=0.8,
  text=white,
  rotate=0.0
]{$T_iT_eT_h$};
\draw (axis cs:2.5,2.5) node[
  scale=0.8,
  text=white,
  rotate=0.0
]{$T_iT_h$};
\draw (axis cs:3.5,2.5) node[
  scale=0.8,
  text=white,
  rotate=0.0
]{$T_iT_eT_h$};
\end{axis}

\end{tikzpicture}
    \caption{Best models for the TT as a function of the amount of training data.}
    \label{fig:heatmap}
    \vspace{-0.5cm}
\end{figure}
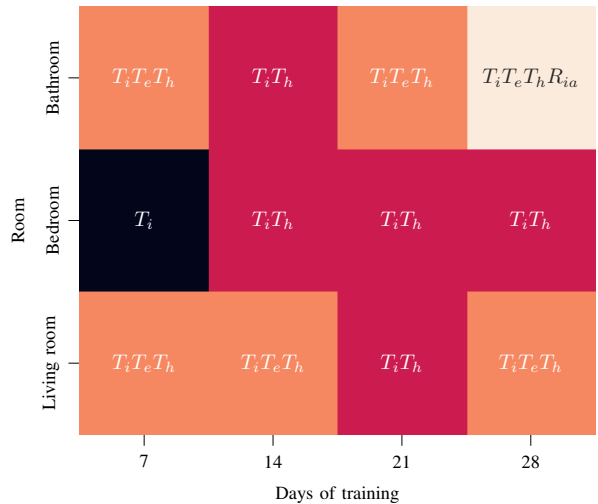

Fig.~\ref{fig:mse} shows the \gls{mse} of the internal temperature predicted by the \gls{tt} for the three rooms we have considered in our study. As expected, the \gls{mse} generally decreases as the available data become richer, but the three rooms all have different trends: while 7 days of data are enough to train the \gls{dt} of the living room, the master bedroom needs significantly more data and has a far higher error.
This can be explained by the fact that the master bedroom is the room which is least occupied during the day: the model is then able to learn about the heating system dynamics only during the night, and the nighttime temperature is kept at a low value of 16$^{\circ}C$. This does not allow it to obtain a sufficiently wide variety of samples to learn from; therefore, the model does not learn as fast as in the other rooms, which show a similar behavior. Nevertheless, all three rooms show large improvements when trained with a longer dataset: the average error for the master bedroom  decreases from more than 1.8$^{\circ}C$ with only 1 day of training to less than 1.2$^{\circ}C$ when trained with a month of data, with roughly a 30\% reduction in the total error. A similar comment can be also made for the other two rooms, where at the beginning of the training the error is around 1.2$^{\circ}C$ and it is reduced to roughly 0.7$^{\circ}C$. We can also consider which model performs best in each room for different amounts of training data: the best model for each considered case is shown in Fig.~\ref{fig:heatmap}, which confirms our analysis. More complex models are usually better with more data, and the evolution of the best model selection follows this trend, with some oscillations.

\begin{figure*}[t]
\centering
\subfloat[Living room.\label{fig:livingroom}]{
\begin{tikzpicture}

\definecolor{darkgray176}{RGB}{176,176,176}

\begin{axis}[
width=\twidth,
height=\theight,
colormap={mymap}{[1pt]
  rgb(0pt)=(1,1,0.8);
  rgb(1pt)=(1,0.929411764705882,0.627450980392157);
  rgb(2pt)=(0.996078431372549,0.850980392156863,0.462745098039216);
  rgb(3pt)=(0.996078431372549,0.698039215686274,0.298039215686275);
  rgb(4pt)=(0.992156862745098,0.552941176470588,0.235294117647059);
  rgb(5pt)=(0.988235294117647,0.305882352941176,0.164705882352941);
  rgb(6pt)=(0.890196078431372,0.101960784313725,0.109803921568627);
  rgb(7pt)=(0.741176470588235,0,0.149019607843137);
  rgb(8pt)=(0.501960784313725,0,0.149019607843137)
},
point meta max=1,
point meta min=0,
tick align=outside,
tick pos=left,
xlabel={Hour},
ylabel={Day},
ytick={0,1,2,3,4,5,6},
yticklabels={Sun,Sat,Fri,Thu,Wed,Tue,Mon},
xtick={0,12,24,36,48,60,72,84},
xticklabels={0,3,6,9,12,15,18,21},
x grid style={darkgray176},
xmin=0, xmax=95,
xtick style={color=black},
y grid style={darkgray176},
ymin=0, ymax=6,
ytick style={color=black}
]
\addplot graphics [includegraphics cmd=\pgfimage,xmin=0, xmax=95, ymin=0, ymax=6] {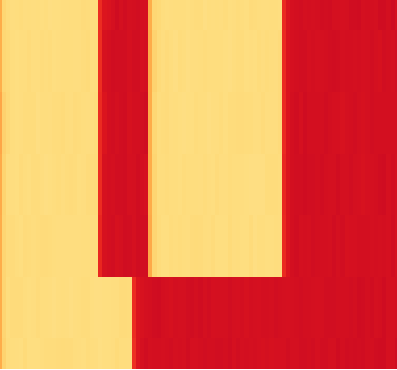};
\end{axis}

\end{tikzpicture}}%
\subfloat[Bathroom.\label{fig:bathroom}]{
\begin{tikzpicture}

\definecolor{darkgray176}{RGB}{176,176,176}

\begin{axis}[
width=\twidth,
height=\theight,
colormap={mymap}{[1pt]
  rgb(0pt)=(1,1,0.8);
  rgb(1pt)=(1,0.929411764705882,0.627450980392157);
  rgb(2pt)=(0.996078431372549,0.850980392156863,0.462745098039216);
  rgb(3pt)=(0.996078431372549,0.698039215686274,0.298039215686275);
  rgb(4pt)=(0.992156862745098,0.552941176470588,0.235294117647059);
  rgb(5pt)=(0.988235294117647,0.305882352941176,0.164705882352941);
  rgb(6pt)=(0.890196078431372,0.101960784313725,0.109803921568627);
  rgb(7pt)=(0.741176470588235,0,0.149019607843137);
  rgb(8pt)=(0.501960784313725,0,0.149019607843137)
},
point meta max=1,
point meta min=0,
tick align=outside,
tick pos=left,
xlabel={Hour},
ylabel={Day},
ytick={0,1,2,3,4,5,6},
yticklabels={Sun,Sat,Fri,Thu,Wed,Tue,Mon},
xtick={0,12,24,36,48,60,72,84},
xticklabels={0,3,6,9,12,15,18,21},
x grid style={darkgray176},
xmin=0, xmax=95,
xtick style={color=black},
y grid style={darkgray176},
ymin=0, ymax=6,
ytick style={color=black}
]
\addplot graphics [includegraphics cmd=\pgfimage,xmin=0, xmax=95, ymin=0, ymax=6] {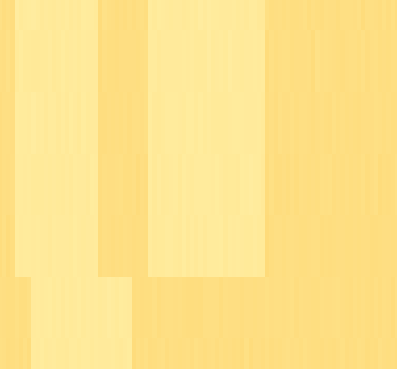};
\end{axis}

\end{tikzpicture}}%
\subfloat[Master bedroom.\label{fig:bedroom}]{
\begin{tikzpicture}

\definecolor{darkgray176}{RGB}{176,176,176}

\begin{axis}[
width=\twidth,
height=\theight,
colorbar,
colorbar style={ylabel={}},
colormap={mymap}{[1pt]
  rgb(0pt)=(1,1,0.8);
  rgb(1pt)=(1,0.929411764705882,0.627450980392157);
  rgb(2pt)=(0.996078431372549,0.850980392156863,0.462745098039216);
  rgb(3pt)=(0.996078431372549,0.698039215686274,0.298039215686275);
  rgb(4pt)=(0.992156862745098,0.552941176470588,0.235294117647059);
  rgb(5pt)=(0.988235294117647,0.305882352941176,0.164705882352941);
  rgb(6pt)=(0.890196078431372,0.101960784313725,0.109803921568627);
  rgb(7pt)=(0.741176470588235,0,0.149019607843137);
  rgb(8pt)=(0.501960784313725,0,0.149019607843137)
},
point meta max=1,
point meta min=0,
tick align=outside,
tick pos=left,
xlabel={Hour},
ylabel={Day},
ytick={0,1,2,3,4,5,6},
yticklabels={Sun,Sat,Fri,Thu,Wed,Tue,Mon},
xtick={0,12,24,36,48,60,72,84},
xticklabels={0,3,6,9,12,15,18,21},
x grid style={darkgray176},
xmin=0, xmax=95,
xtick style={color=black},
y grid style={darkgray176},
ymin=0, ymax=6,
ytick style={color=black}
]
\addplot graphics [includegraphics cmd=\pgfimage,xmin=0, xmax=95, ymin=0, ymax=6] {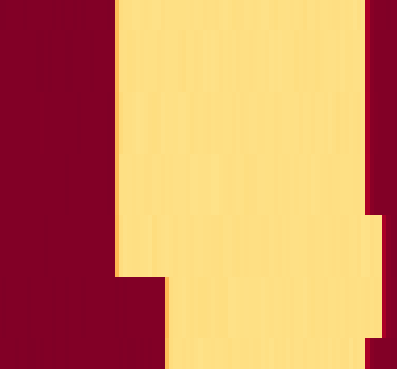};
\end{axis}

\end{tikzpicture}}%
    \caption{Heatmap of the probability for each room to be occupied over time in the synthetic \gls{rt} model.}
\label{fig:rt_prob}
\vspace{-0.5cm}
\end{figure*}

There are three main takeaways from this analysis: \textit{(i)}, personalized models have to be created and trained for each room, even in the same house, as the heating and temperature dynamics vary greatly between rooms, depending on their size, isolation, exposure to sunlight, and heating units; \textit{(ii)}, the larger the training set, the better the results, as it is possible to observe a wider variety of samples; \textit{(iii)}, a \gls{dt} can effectively help to understand when a model is ready to be deployed, since it can be trained on a portion of data and tested with fresh data, and then used whenever the error falls under a scenario-specific threshold, and \emph{iv}, the \gls{dt} can include different types of models with different levels of complexity, which can be hot-swapped as new data become available and richer models can be trained effectively. For a wider review of RC-equivalent model parameters and training, we refer the reader to~\cite{leprince2022fifty}.

On the other hand, the smart home deployment did not provide room-level occupant data, so we assumed a simple Markovian model. For each room, we considered a maximum number of occupants $N_{\max}$, each of whom behaves independently. The occupant is in state 1 if they are inside the room, and the evolution of their state is driven by the matrix $\mathbf{P}(d,h)$, which is a function of the day of the week and of the current time. Fig.~\ref{fig:rt_prob} shows the average probability of each room being occupied over the whole week: the bedroom is occupied during the night, with a slightly shifted timeframe in the weekend, while the bathroom and living room are usually occupied during the day and evening, with different transition probabilities.  Naturally, the actual probability of a room being occupied at a given time will depend on the state, but we can see that the bedroom is more predictable, while, as expected, the bathroom has a much more random pattern throughout the day. The assumed probability for someone to be in the bathroom is low but constant throughout the day, while the occupants remain in the living room for longer times during the morning and evening, resulting in a higher probability of the room being occupied. The three rooms also have different numbers of occupants: we assume that the whole family of 4 may be present in the living room, while only 2 people use the master bedroom, and only one person at a time can be in the bathroom (in this case, one person leaving the bathroom and another entering would count as one person remaining in the room).

The use of an artificial model for the room occupancy reduces its accuracy in the real deployment, but the architecture can easily accommodate new sensors and \glspl{dt}, automatically retraining the \gls{drl} agent after updates to the underlying models. In fact, since a time-dependent \gls{rt} would need to be trained over multiple weeks to figure out the family's habits, starting from a relatively accurate prior would significantly reduce the \gls{dt} training time.

\subsection{Agent evaluation}
In this section, we evaluate the performance of the agent, which leverages the \gls{rt} and \gls{tt} to forecast actions which improve the overall status of the system.

\begin{table}[b]
	\vspace{-0.3cm}
	\centering
	\scriptsize
	\caption{MDP parameters.}
	\label{tab:mdp_param}
	\begin{tabular}{lcc}
		\toprule
        Parameter & Symbol & Value\\
		\midrule
        Episode duration (steps) & $L$ & 1000\\
        Step duration (minutes) & $t$ & 15\\
        Comfort temperature & $T_0$ & $18^\circ C$\\
        Threshold temperature & $T_{\text{lim}}$ & $3^\circ C$\\
        Energy cost & $\alpha$ & $0.25$\\
        Swing penalty & $\beta$ & $0.2$\\
        Discount factor & $\lambda$ & 0.95\\
		\bottomrule
	\end{tabular}
\end{table}

The temperature control \gls{mdp} is relatively simple: the state of the system is represented by the states of the two twins, which include information about the thermal state of the room and heating system, as well as the date and current occupancy of the room. The possible actions available to the agent are $A$ levels of heating system power, which are normalized from 0 (the heating system is off) to 1, corresponding to the maximum available heating power. The reward function for taking action $a$ when there are $o$ people in the room, resulting in an internal temperature $T_i$, is defined as follows:
\begin{equation}
\begin{aligned}
R(T_i,o,a)=\mathbbm{1}(T_i-T_0)\mathbbm{1}(o-1)-\alpha a\\
-\beta\mathbbm{1}(|T_i-T_0|-T_{\text{lim}})\left(1-2\mathbbm{1}(a)\right)\text{sign}(T_i-T_0),
\end{aligned}
\end{equation}
where $T_0$ and $T_{\text{lim}}$ are user-defined threshold temperatures, $\mathbbm{1}(\cdot)$ is the stepwise function, equal to 1 if the argument is greater than 0 and 0 otherwise, and $\alpha$ and $\beta$ are penalty parameters representing the cost of energy and a generic penalty for . In other words, the agent receives a reward 1 if the temperature is above the comfort level $T_0$, but only if there are people in the room; on the other hand, using energy to heat always has a cost. The second penalty term is used to avoid excessive temperature swings: if the room is too warm or too cold by more than $T_{\text{lim}}$ degrees, actions that do not lead the system back toward equilibrium (i.e., not turning on the heat if it is too cold, or turning it on when the room is already too hot) are penalized. The values we selected for the parameters are listed in Table~\ref{tab:mdp_param}.

We then implement a \gls{dqn} model, with an architecture that depends on the \gls{dt} models, and is defined in Table~\ref{tab:dqn_param}, along with the main training parameters: the \gls{rt} only needs $N_{\text{RT}}=3$ parameters (the day of the week, time, and number of current occupants of the room), while the \gls{tt} can have a variable $N_{\text{TT}}$ depending on the RC-equivalent model. The $T_i$ model only has a single parameter, while the more complex $T_iT_eT_hR_{ia}$ model has 4. The final input value is the ambient temperature $T_a$ outside the room. The neural network uses a Softmax exploration policy, whose temperature decreases logarithmically from 1 to $10^{-6}$, and the Adam optimizer, with ReLU activation.

\begin{table}[b]
	\vspace{-0.3cm}
	\centering
	\scriptsize
	\caption{DQN architecture and training parameters.}
	\label{tab:dqn_param}
	\begin{tabular}{lcc}
		\toprule
        \textbf{Layer} & \textbf{Symbol} & \textbf{Parameters}\\
		\midrule
        Input & $L_0$ &$N_{\text{TT}}+N_{\text{RT}}+1$\\
        Input & $L_1$ & $10L_0$\\
        Input & $L_2$ & $5L_1$\\
        Output & $O$ & $A$\\
        \midrule
        \textbf{Hyperparameter} & \textbf{Symbol} & \textbf{Value}\\
		\midrule
        Activation function & $\phi_a$ & ReLU\\
        Learning rate & $\ell$ & $10^{-5}$\\
        Dropout probability & $p_d$ & 0.1\\
        Epochs & $E$ & 25\\
        Episodes per epoch & $e$ & 20\\
		\bottomrule
	\end{tabular}
\end{table}

\begin{figure}
\centering
\begin{tikzpicture}

\definecolor{darkgray176}{RGB}{176,176,176}
\definecolor{darkslategray66}{RGB}{66,66,66}
\definecolor{lightgray204}{RGB}{204,204,204}

\begin{axis}[
width=\fwidth,
height=\fheight,
legend cell align={left},
legend style={
  fill opacity=0.8,
  draw opacity=1,
  text opacity=1,
  legend columns=2,
  at={(0.97,0.03)},
  anchor=south east,
  draw=lightgray204
},
tick align=outside,
tick pos=left,
unbounded coords=jump,
x grid style={darkgray176},
xmin=-0.5, xmax=3.5,
xtick style={color=black},
xtick={0,1,2,3},
xticklabels={7,14,21,28},
y grid style={darkgray176},
ylabel={Reward},
ymin=-0.55, ymax=0.55,
ytick style={color=black}
]
\draw[draw=none,fill=color1] (axis cs:-0.4,0) rectangle (axis cs:-0.133333333333333,0.125395139892399);
\addlegendimage{ybar,ybar legend,draw=none,fill=color1}
\addlegendentry{Agent (bathroom)}

\addplot[dashed,semithick,color1]
table{
-10 0.1855
40 0.1855
};
\addlegendentry{Ideal (bathroom)}

\draw[draw=none,fill=color1] (axis cs:0.6,0) rectangle (axis cs:0.866666666666667,0.167805725699663);
\draw[draw=none,fill=color1] (axis cs:1.6,0) rectangle (axis cs:1.86666666666667,0.170443174745142);
\draw[draw=none,fill=color1] (axis cs:2.6,0) rectangle (axis cs:2.86666666666667,0.1504280849576);
\draw[draw=none,fill=color2] (axis cs:-0.133333333333333,0) rectangle (axis cs:0.133333333333333,-0.306024611233175);
\addlegendimage{ybar,ybar legend,draw=none,fill=color2}
\addlegendentry{Agent (bedroom)}

\addplot[dashed,semithick,color2]
table{
-10 0.5237
40 0.5237
};
\addlegendentry{Ideal (bedroom)}

\draw[draw=none,fill=color2] (axis cs:0.866666666666667,0) rectangle (axis cs:1.13333333333333,-0.10215952398181);
\draw[draw=none,fill=color2] (axis cs:1.86666666666667,0) rectangle (axis cs:2.13333333333333,0.324597950831056);
\draw[draw=none,fill=color2] (axis cs:2.86666666666667,0) rectangle (axis cs:3.13333333333333,0.320778125);
\draw[draw=none,fill=color3] (axis cs:0.133333333333333,0) rectangle (axis cs:0.4,-0.278480551859736);
\addlegendimage{ybar,ybar legend,draw=none,fill=color3}
\addlegendentry{Agent (living room)}

\addplot[dashed,semithick,color3]
table{
-10 0.492
40 0.492
};
\addlegendentry{Ideal (living room)}

\draw[draw=none,fill=color3] (axis cs:1.13333333333333,0) rectangle (axis cs:1.4,0.262259375);
\draw[draw=none,fill=color3] (axis cs:2.13333333333333,0) rectangle (axis cs:2.4,0.29685625);
\draw[draw=none,fill=color3] (axis cs:3.13333333333333,0) rectangle (axis cs:3.4,0.287271875);
\end{axis}

\end{tikzpicture}
    \caption{\gls{drl} agent performance at convergence as a function of the number of days of available data.}
    \label{fig:rl_perfect_rt}
        \vspace{-0.5cm}
\end{figure}
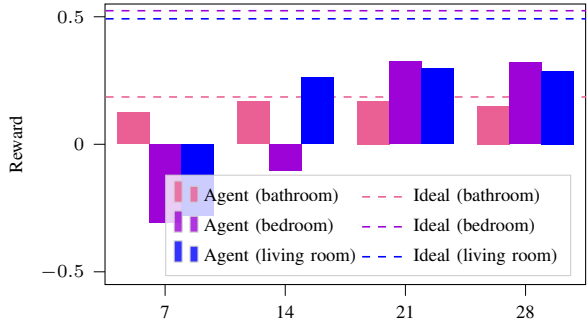

Fig.~\ref{fig:rl_perfect_rt} shows the performance of the \gls{drl} agent, as measured by the reward, when it is trained using a \gls{tt} with a limited amount of data. In all cases, the \gls{rt} is assumed to be perfect. We can immediately note that the bedroom needs at least 3 weeks of data to control the room temperature properly, while the bathroom can achieve good performance with a single week. This is consistent with the \gls{mse} performance of the models: while the bathroom dynamics are easy to learn, and the model predictions are similar to reality, the significant errors in the bedroom mean that the virtual training leads the \gls{drl} agent to the wrong policy, with catastrophic results. As the \gls{tt} improves, so does the performance of the agents, gradually reaching the optimum. We also highlight that the optimal performance for a room depends on how much it is occupied: the dashed lines in the figure show the performance of an ideal system with perfect room occupancy prediction and free energy, i.e., the upper bound to real performance for each room. The upper bound is extremely hard to reach in the bigger rooms, as the real agent has to deal with winter temperatures outside, and consequently frequently uses energy to power the heating system. On the other hand, the bathroom does not directly face any walls, and maintaining a comfortable temperature is relatively easy, so it is possible to achieve an almost perfect performance.

\begin{figure}
\centering
    \input{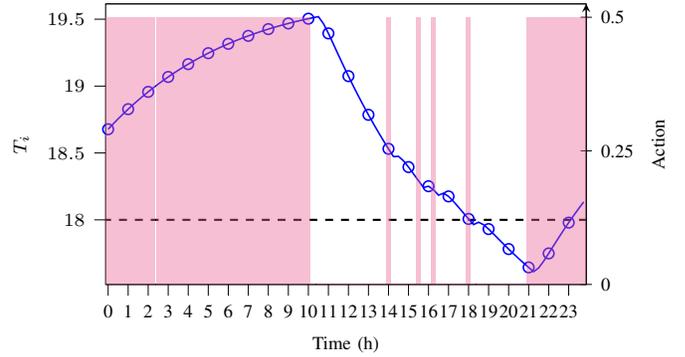}
    \caption{Temperature evolution and control actions in the master bedroom over the course of a day.}
    \label{fig:daytemp}
    \vspace{-0.5cm}
\end{figure}
To provide a better understanding of how the model works, we also show an example sequence of actions performed by the \gls{drl} agent over the course of a day for the master bedroom in Fig.~\ref{fig:daytemp}. The room is only occupied at night, although during the day has a non-zero probability of having occupants. During the night (i.e., leftmost part of the plot), the agent heats up the room beyond the minimum desired temperature of 18$^{\circ}C$, with the heating system at half power. In the morning, the probability of having at least one occupant significantly drops, hence the \gls{drl} agent stops to heat the room. Sporadic adjustments are made through the day, since the \gls{drl} agent wants to keep the room above the minimum temperature to avoid any reward losses. In the evening the \gls{drl} agent starts to heat the room again some time before the usual bedtime, raising the temperature above the minimum threshold in time for them to go to bed. Obviously, having a perfect representation of the occupancy would avoid the \gls{drl} agent to sporadically heat the room when the probability of having occupants is low. Still, Fig.~\ref{fig:daytemp} confirms that the system actually learns a non-obvious policy from the two digital twins, and improves the system overall.

\subsection{Automatic Agent Definition}
In the use case we defined, the \gls{drl} agent is still partially hand-designed: the selection of the \gls{dt} models was performed manually, as well as the architecture and hyperparameter search for the training. However, the \gls{wot} architecture we define provides significant room for expansion and for automation, as agents, models, sensors, and actuators are all controlled by the same platform.

We can foresee a use case in which multiple twins (e.g., including multiple rooms, sensors, and activities) can be automatically mixed and matched to tasks (as represented by actuators). The full potential of the proposed architecture is to automate \gls{dqn} training as well, trying different state definitions and combinations of actions, as well as automatically optimizing hyperparameters for each agent. In this case, the only hand-designed part of the system would be the reward function, which would be ideal in a domotic scenario: the inhabitants only need to give the smart home a simple list of priorities, which they can adjust at any time, and the architecture will figure out which agents it will need to deploy, which twins they will need to accomplish their tasks, and which tasks are functionally independent and can be separated to simplify the action space.

\section{Conclusions}
\label{sec:conclusions}
In this paper, we have a presented a novel framework to realize digital twins based on the Web of Thing standard. Our architecture allows to build customized digital twins for complex scenarios, leveraging data from custom made sensors or legacy ones, by integrating it with already existing platforms. Our results indicate the benefits of using \glspl{dt} for model-based reinforcement control, as they allow to test different models and deploy the one which fits the scenario the best. This has been confirmed by deploying our platform on a real scenario on which a legacy system was already running, and integrating our platform with the existing one.

Future work on this topic will be focused on the extension of our framework to other sensor types in a smart home, a more in-depth analysis of the room twin, and extended results on other house types and rooms, which will better generalize our findings. Furthermore, the proposed automation of agent selection and \gls{mdp} design is an interesting future direction.

\bibliographystyle{IEEEtran}
\bibliography{./biblio} 

\end{document}